# High resolution meta-stereomicroscope based on birefringent meta-optics

LIHENG YAN,[1] HAOWEN LIANG,[1,2,7*] EMILIANO R. MARTINS,[3,†] YIKUN LIU,[4] TAO LI,[5] THOMAS F. KRAUSS,[6] JUNTAO LI,[1,2,8*] AND XUE-HUA WANG[1,2]

[1] *State Key Laboratory of Optoelectronic Materials and Technologies, School of Physics, Sun Yat-Sen University, Guangzhou 510275, China.*
[2] *Quantum Science Center of Guangdong-Hong Kong-Macao Greater Bay Area (Guangdong), Shenzhen, China.*
[3] *São Carlos School of Engineering, Department of Electrical and Computer Engineering, University of São Paulo, 13566-590, Brazil.*
[4] *Guangdong Provincial Key Laboratory of Quantum Metrology and Sensing, School of Physics and Astronomy, Sun Yat-Sen University, Zhuhai 519080, China.*
[5] *National Laboratory of Solid State Microstructures, Key Laboratory of Intelligent Optical Sensing and Manipulation, Jiangsu Key Laboratory of Artificial Functional Materials, Collaborative Innovation Center of Advanced Microstructures, College of Engineering and Applied Sciences, Nanjing University, Nanjing.*
[6] *School of Physics, Engineering and Technology, University of York, York, YO10 5DD, UK.*
[†] *Deceased on 15 July 2026.*
*[7] *e-mail: lianghw26@mail.sysu.edu.cn*
*[8] *e-mail: lijt3@mail.sysu.edu.cn*

**Abstract:** Achieving both high lateral and high depth resolution is a longstanding goal in stereomicroscopy. Although meta-optics have revolutionized lens design by alleviating the physical constraints of conventional optical architectures, existing metalens-assisted stereomicroscopes still suffer from field of view (FOV) mismatch between meta-optical and conventional optical components in stereomicroscopes, thereby limiting their imaging performance. Here, we show that this mismatch can be fundamentally eliminated through a fully meta-optical architecture. The integrated system enables flexible control of numerical aperture and magnification with larger depth of field (DOF) and FOV than those of the metalens-assisted stereomicroscopes. Experimentally, we achieve an integrated meta-stereomicroscope with a lateral resolution of 435 nm, surpassing the performance of previously reported stereomicroscopes. Empowered by a stereo neural network, the system enables straightforward reconstruction of high-resolution three-dimensional surface morphology with a depth resolution of 1026 nm, demonstrating the capability to simultaneously achieve high lateral and depth resolution imaging. This integrated architecture operates in both transmission and reflection modes for biomedical imaging and industrial inspection, highlighting its broad applicability for real-time observation across biomedical and industrial scenarios.

## 1 Introduction

With the rapid development of the life sciences and the semiconductor industry, microscopic observation has increasingly evolved from two-dimensional (2D) imaging to real-time three-dimensional (3D) imaging. This transition imposes significantly greater demands on both lateral and depth resolution than 2D imaging **[1–3]**. Therefore, several 3D microscopic techniques, including confocal microscopy **[4]**, digital holographic microscopy **[5]**, quantitative phase microscopy **[6]**, white-light interferometry **[7]**, and optical diffraction tomography **[8]**, can meet the high-resolution requirement. However, many of these techniques rely on axial scanning to reconstruct the depth information over several micrometers, limiting the imaging speed. In contrast, stereomicroscopes enable single-shot, real-time 3D imaging with a relatively simple optical configuration and have been widely adopted in various applications **[9]**. To accommodate human binocular vision **[10, 11]**, commercial stereomicroscopes employ dual independent optical paths, in which left- and right-view images are captured by two front-end objective lenses. However, unlike macroscopic stereoscopic imaging systems **[12–14]**, the physical separation required between the two objective lenses fundamentally limits the achievable numerical aperture (NA) and thus the imaging resolution of stereomicroscopes **[15]**. More generally, existing 3D microscopic imaging systems based on conventional optical components remain bulky and complex, hindering their integration into compact imaging platforms **[16–18]**.

Meta-optics **[19–21]** exploits ultrathin meta-atoms to manipulate the amplitude, phase, and polarization of electromagnetic fields, enabling multiple functionalities to be integrated into a single device. Owing to their powerful wavefront-engineering capabilities, metalenses have been applied to many microscopy systems, such as super-resolution imaging **[22]** and wide-field microscopy **[23, 24]**. Their potential for 3D microscopic imaging, however, has only recently been recognized **[25–30]**. Recently, a birefringent metalens is shown to enhance the NA of a Greenough-type stereomicroscope to 0.4 **[15]** by first forming an intermediate image of the sample with the metalens and then re-imaged by the stereomicroscope. However, this configuration still relies on a conventional stereomicroscope architecture, resulting in relatively large system dimensions. In the metalens-assisted stereomicroscope, the secondary imaging of the metalens by the stereomicroscope causes the overall depth of field (DOF) to be constrained by the stereomicroscope. More importantly, it suffers from a fundamental field of view (FOV) mismatch between the metalens and the stereomicroscope. This mismatch limits the FOV of this system (see the Discussion section for a detailed analysis), restricting its practical applicability.

Here, we demonstrate a fully meta-optical architecture that eliminates the FOV mismatch of the metalens-assisted stereomicroscope **[15]**. The key component is a single birefringent metalens that encodes two distinct phase profiles, enabling the simultaneous acquisition of left- and right-view images through orthogonal polarization states (see **Fig. 1(a)**). By eliminating the need for conventional optics, this single device enables flexible control of NA and magnification with larger DOF and FOV than those of the metalens-assisted stereomicroscope. Experimentally, the metalens with an NA of 0.6 is integrated with a polarization imaging sensor to form a compact imaging module with a size of 40 mm × 40 mm × 40 mm (see **Fig. 1(c)**). By further integrating an illumination system and a sample stage, a meta-stereomicroscope with a size of 110 mm × 110 mm × 140 mm is realized (see **Fig. 1(b)**). Operating at a wavelength of 532 nm, the system achieves a lateral resolution of 435 nm and enables real-time 3D microscopic imaging of biological samples. This performance surpasses previously reported stereomicroscopes in both system compactness and imaging resolution. Combined with a stereo neural network, it further enables high-resolution 3D surface morphological reconstruction with a depth resolution of 1026 nm, demonstrating the capability to simultaneously achieve high lateral and depth resolution. This integrated architecture operates in both transmission and reflection modes, highlighting its broad potential for applications ranging from biomedical imaging to industrial inspection.

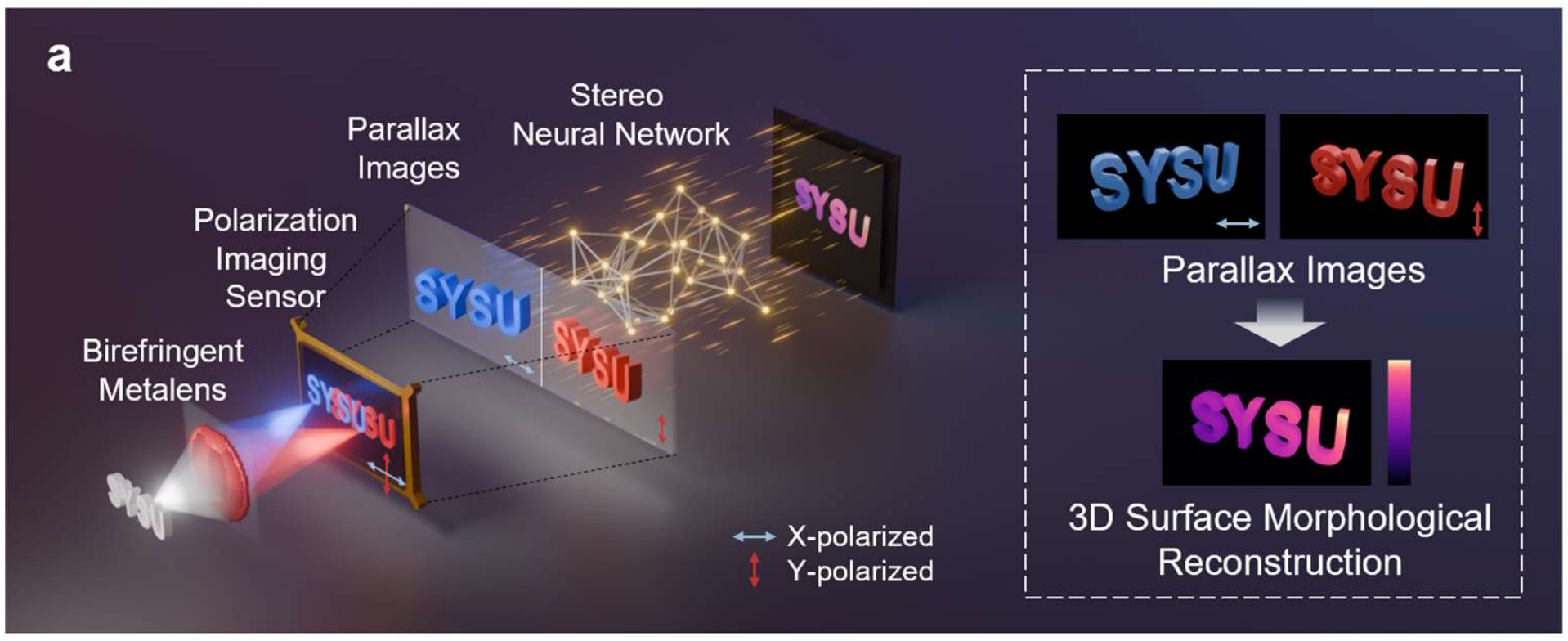

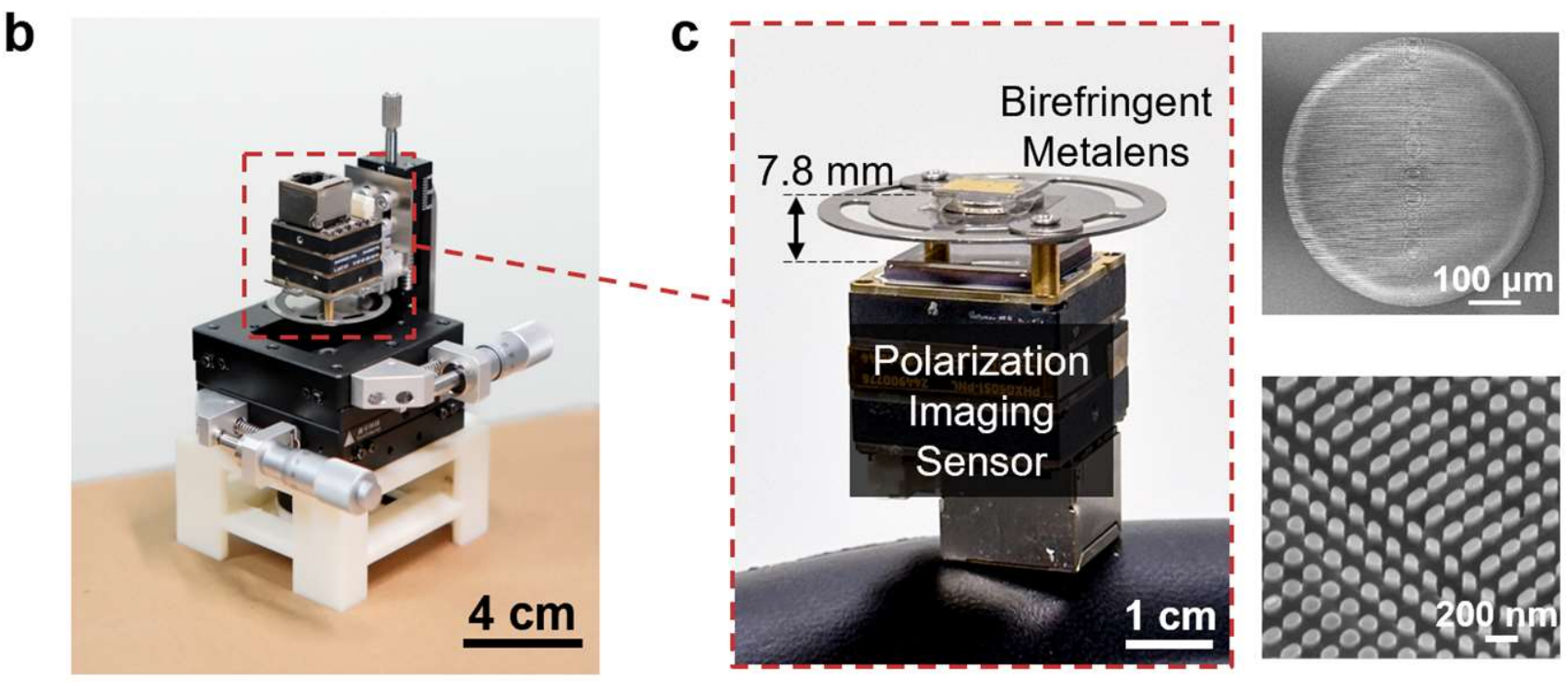

**Fig. 1. Schematics and photographs of the meta-stereomicroscope. (a)** Working principle of the meta-stereomicroscope. **(b)** Photograph of the meta-stereomicroscope, with the imaging module enclosed by the red dashed box. **(c)** Magnified view of the imaging module of the meta-stereomicroscope with an inset showing the scanning electron microscope (SEM) image of the birefringent metalens.

## 2 Results

### Design, Fabrication, and Characterization of the Birefringent Metalens

**Fig. 2(a)** schematically illustrates the design of the birefringent metalens for the meta-stereomicroscope, where the blue and red colors denote the light paths corresponding to the two distinct phase profiles. These phase profiles steer the incident light along two tilted optical axes: the left optical path corresponds to x-polarized light, whereas the right optical path corresponds to y-polarized light. This configuration generates two perspectives of the same object, enabling 3D imaging through the parallax effect **[10, 11]**. For each channel, the optical axis remains tilted with the same angle on the image side, thereby minimizing spherical aberration and maintaining high imaging quality. The metalens is designed to operate under water immersion at a wavelength of 532 nm with an NA of 0.6. This water-immersion configuration not only facilitates the observation of living cells in aqueous environments but also enables a higher NA than would be achievable under air immersion.

We introduce an aplanatic phase profile based on the generalized laws of refraction and imaging **[31]**. Accordingly, each polarization state is encoded with an aspherical phase profile, as described below and illustrated in **Fig. 2(b)**:

$$\varphi_l(x,y)=\frac{2\pi n_{object}}{\lambda_0}\left(\frac{s}{\cos\theta}-\sqrt{x^2+y^2+s^2}\right)+\frac{2\pi n_{image}}{\lambda_0}\left(\frac{v}{\cos\theta}-\sqrt{(x+L)^2+y^2+v^2}\right) \quad (1)$$

$$\varphi_r(x,y)=\frac{2\pi n_{object}}{\lambda_0}\left(\frac{s}{\cos\theta}-\sqrt{x^2+y^2+s^2}\right)+\frac{2\pi n_{image}}{\lambda_0}\left(\frac{v}{\cos\theta}-\sqrt{(x-L)^2+y^2+v^2}\right) \quad (2)$$

where $x$ and $y$ are the transverse coordinates; $s$ and $v$ are the object and image distances, respectively; and $\theta$ is the observation angle, set to be 6° to match the visual angle of human eyes. $\lambda_0$ = 532 nm is the wavelength in air, and $n_{object}$ and $n_{image}$ are the refractive indices of the object and image spaces, respectively. The metalens is designed with a magnification of 8, and $L=(s+v)\cdot\tan\theta$ is the separation distance between the left and right images on the sensor plane. The NA is defined as NA=$n_{water}\,sin\,\alpha$, where $\alpha$ (see **Fig. 2(a)**) is the maximum acceptance half-angle with respect to the optical axis. The refractive index of water is $n_{water}$=1.33.

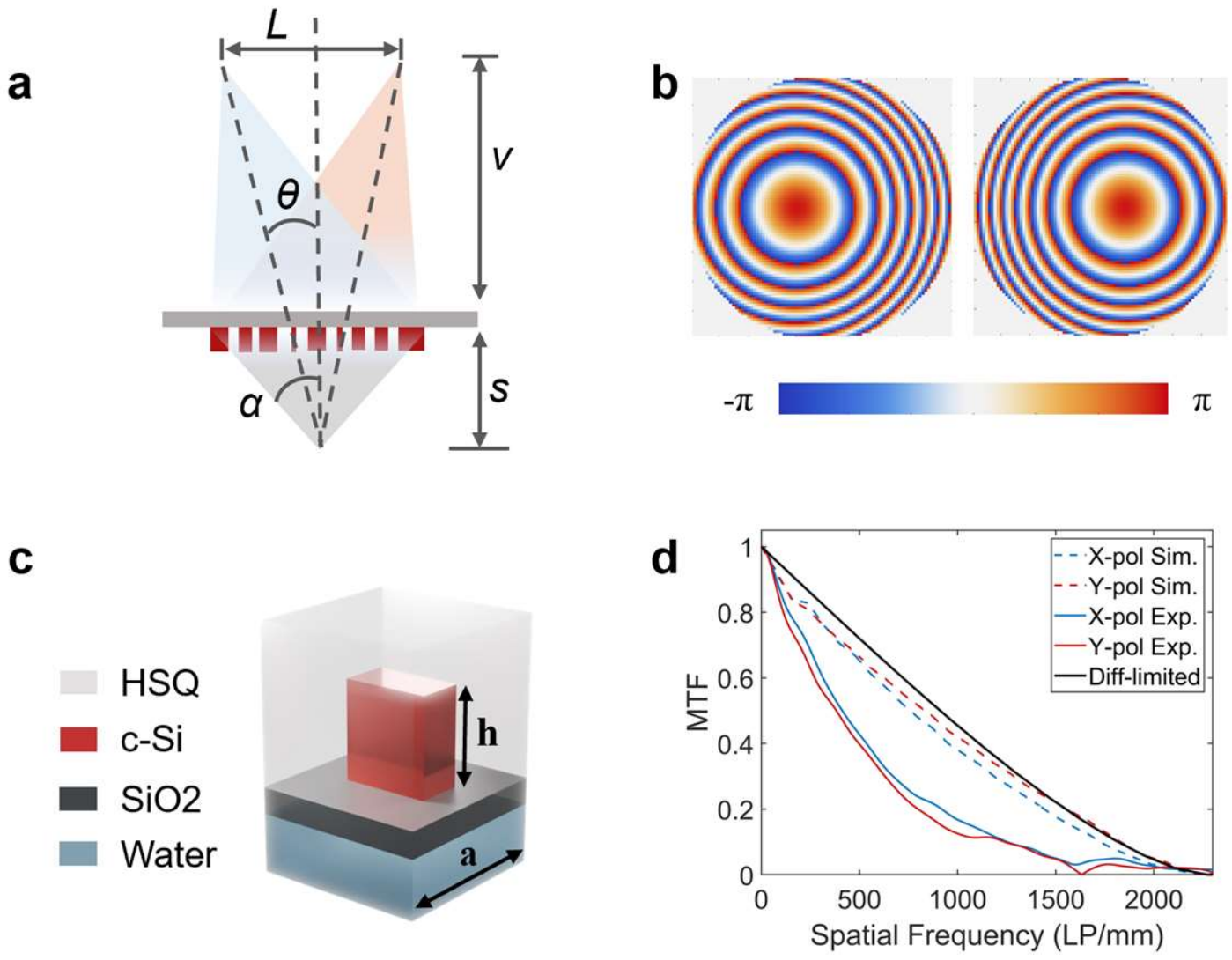

**Fig. 2. Design and performance of the birefringent metalens.** (**a**) Optical axes and optical paths for stereoscopic imaging. The blue and red paths represent x- and y-polarized light, respectively. (**b**) Ideal phase profile of the demonstrated birefringent metalens. The left and right figures correspond to the phase distributions for x- and y-polarized light. (**c**) Schematic of a meta-atom. (**d**) Simulated and experimentally measured modulation transfer function (MTF) of the birefringent metalens for orthogonal polarizations. The black solid curve corresponds to the diffraction-limited MTF of a conventional lens with NA = 0.6.

The metalens employs rectangular meta-atoms (see **Fig. 2(c)**) to encode two different phase profiles into orthogonal polarization states. To protect the metalens from damage and contamination, the nanopillars are encapsulated between hydrogen silsesquioxane (HSQ) and silica layers. Crystalline silicon (c-Si) is chosen as the constituent material for the nanopillars due to its high refractive index contrast relative to the surrounding medium. The nanopillars have a height of 220 nm, and the period of the meta-atom is 170 nm. The lengths and widths of the nanopillars vary from 50 nm to 140 nm (see **Materials and Methods** and **Section S1 of the Supplemental Information** for details of phase and transmission responses). Notably, the meta-atoms here only exhibit polarization-dependent phase responses, rather than functioning as polarization converters **[32, 33]**. The focusing efficiencies of the left and right optical paths are simulated to be 47% and 44%, respectively, using a commercial finite-difference time-domain (FDTD) solver (ANSYS, Inc.) (see **Section S2, Supplemental Information** for calculation details). The focusing efficiency is defined as the ratio of the optical power enclosed within an area three times the full width at half maximum (FWHM) of the focal spot to the total incident power on the metalens **[14]**. The moderate focusing efficiency arises from the polarization-multiplexed design, which makes it challenging for the meta-atoms to simultaneously provide the required phase responses and maintain high transmittance. Further improvements in efficiency could be achieved by employing non-polarization-multiplexed designs or lower-loss materials **[20]**.

The theoretical lateral resolution of the metalens is evaluated from its MTFs, which are calculated from the simulated point spread functions (PSFs) shown in **Fig. S2(b)**. The resulting MTFs for both x- and y-polarized illumination are presented as dashed lines in **Fig. 2(d).** Using an MTF threshold of 0.1 **[34]**, the simulated cutoff frequencies are 1797 lp/mm and 1914 lp/mm for x- and y-polarized light, respectively. These results indicate that the metalens provides nearly identical imaging performance for both polarization states and achieves a resolution close to the diffraction limit of a conventional lens with NA = 0.6.

Based on the above design, we fabricate a birefringent metalens with a diameter of 500 μm and a focal length of $f$ = 330 μm using electron-beam lithography (EBL) (see **Methods** for details), as shown in the inset in **Fig. 1(c)**. The focusing efficiencies of the left and right optical paths are measured to be 37% and 33%, respectively, using the optical setup shown in **Fig. S2(c)**. The measured PSFs are presented in **Fig. S2(d)**, from which a DOF of 8.1 μm is obtained (see **Fig. S2(e)**). This DOF is comparable to the depth measurement range of stereomicroscopy. The corresponding MTFs, calculated from the measured PSFs, are shown as solid lines in **Fig. 2(d)**. At an MTF threshold of 0.1, the cutoff frequencies are 1236 lp/mm and 1233 lp/mm for x- and y-polarized light, respectively.

### Meta-Stereomicroscope in Transmission Mode

We evaluate the performance of the meta-stereomicroscope under transmission-mode illumination, as shown in **Fig. S3(a)** (see **Materials and Methods** for details). The inclined-optical-axis phase profile, designed according to stereoscopic imaging geometry, separates the left and right optical paths. However, in a single-metalens imaging configuration, the limited spatial separation between the two images on the sensor can lead to overlap and cross-talk. To overcome this issue, a polarization imaging sensor (Lucid Vision Labs, PHX050S1-P/Q) equipped with on-chip polarizers is used to separate and simultaneously record the x- and y-polarized images. The maximum resolvable spatial frequency of the sensor is 72 lp/mm. Additional details of the polarization imaging sensor are provided in **Section S3 of the Supplemental Information**.

To prevent the pixel size of the polarization imaging sensor from limiting the imaging resolution, $s$ and $v$ are adjusted to 0.45 mm and 7.8 mm, corresponding to a magnification of 23 with object space immersed in water. Although a higher magnification can further mitigate the pixel-sampling limitation, it reduces the achievable FOV for a sensor with a finite area. Under these conditions, the sensor imposes an upper limit of 1665 lp/mm on the maximum resolvable spatial frequency at the object plane. As a result, the imaging module of the meta-stereomicroscope, consisting of the metalens and the sensor, achieves a compact form factor of only 40 mm × 40 mm × 40 mm (see **Fig. 1(c)**). The sample-to-sensor distance, defined as the distance between the object plane and the sensor plane, is approximately 8.3 mm. To enhance image resolution and contrast, an infinity-corrected objective lens (20×, NA = 0.4) is placed near the light-emitting diode (LED) to provide focused illumination **[35]**. With further integration of this illumination system and a sample stage, we realize a compact meta-stereomicroscope with dimensions of 110 mm × 110 mm × 140 mm (see **Fig. 1(b)**).

To evaluate the optical resolution, a 1951 United States Air Force (USAF) resolution target is used as the test object. Images corresponding to the left and right optical paths are simultaneously captured by the polarization imaging sensor, as shown in **Fig. 3(a) and (b)**. Features in Group 10 Element 2, corresponding to a spatial frequency of 1149 lp/mm, are clearly resolved in both channels. This corresponds to a half-pitch resolution of 435 nm at a wavelength of 532 nm. A 43 μm-diameter FOV is maintained at this lateral resolution (see **Fig. S4)**. For comparison, **Fig. 3(c)** shows an image of the same region acquired using a commercial Greenough-type stereomicroscope (Yong Heng, XTZ-05T, NA = 0.15) for its highest resolution setup. Owing to its low NA, the corresponding features are completely unresolved. **Fig. 3(d)** further presents an image of the same region acquired using the metalens-assisted stereomicroscope (setup similar to [15]) operated at the same total magnification of 23× as the meta-stereomicroscope. Although it can achieve the same 435-nm resolution, the FOV is restricted by the FOV mismatch between the metalens and the commercial stereomicroscope, as explained in detail in the Discussion section. These results highlight the substantial advantage of the meta-stereomicroscope introduced here.

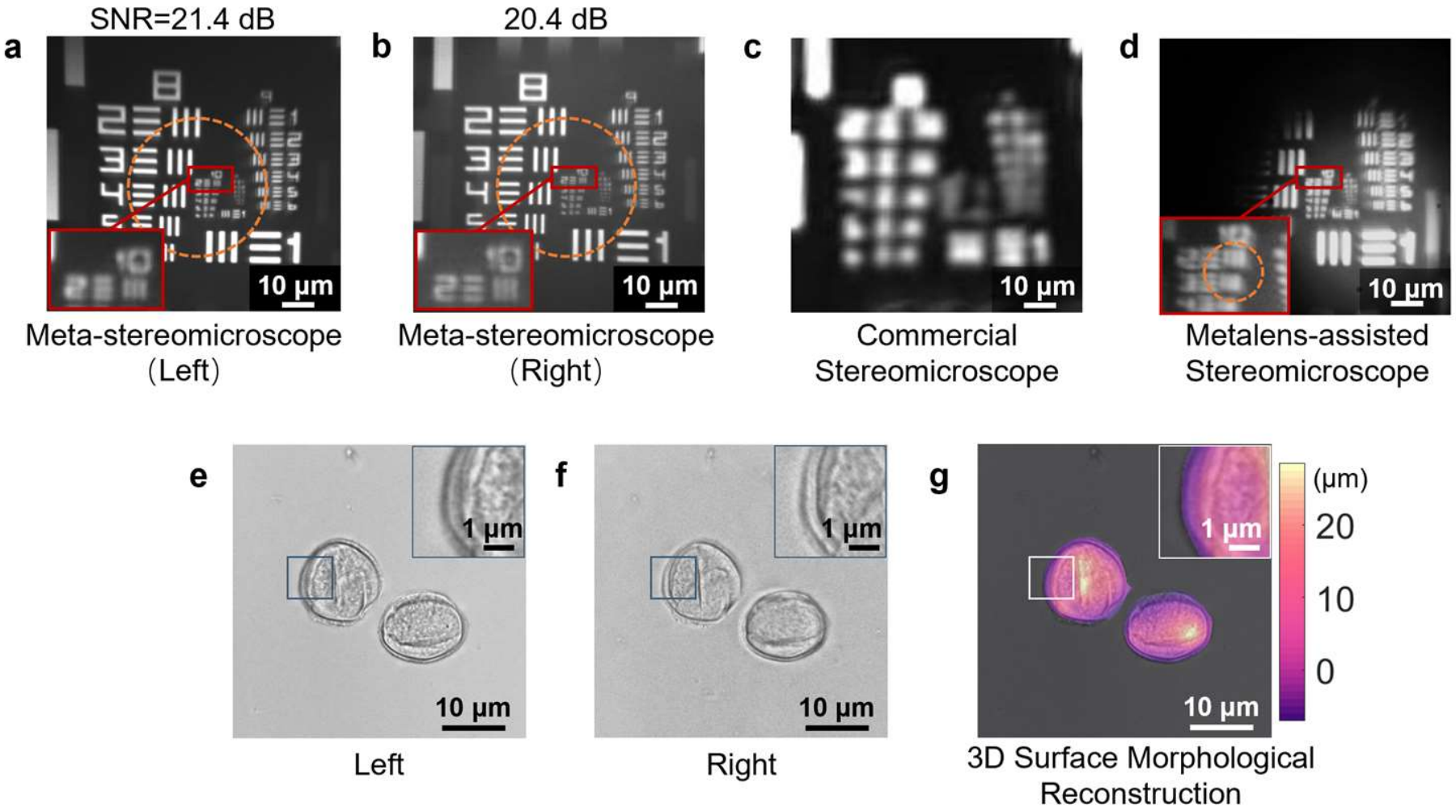

**Fig. 3. Imaging performance of the meta-stereomicroscope under transmission-mode illumination.** (**a–d**) Imaging of the 1951 USAF resolution target using (**a**) the left optical path for x-polarized light, (**b**) the right optical path for y-polarized light, (**c**) a commercial stereomicroscope and (**d**) a metalens-assisted stereomicroscope. Insets in (**a**), (**b**) and (**d**) show magnified views of Group 10, Element 2, and the orange circles indicate the FOV boundaries. (**e–g**) Imaging results of pollen grains from (**e**) the left and (**f**) the right optical paths. (**g**) 3D surface morphological reconstruction merged with image (**e**), showing simultaneous high lateral and depth resolution imaging capability.

To quantify the cross-talk suppression, the signal-to-noise ratio (SNR) is defined as SNR=10×lg ($I_{signal}/I_{noise}$), where $I_{signal}$ and $I_{noise}$ refer to the intensity of the clearly resolvable signal and the standard deviation of the background noise intensity, respectively **[24]**. A non-polarization imaging sensor (Imaging Source, DMM27UJ003-ML) is used for comparison, where the cross-talk is treated as noise. The average SNRs obtained by the non-polarization imaging sensor and the polarization imaging sensor are 13.0/21.4 dB for the left optical path and 12.9/20.4 dB for the right optical path, demonstrating effective cross-talk suppression (see **Fig. S5)**.

To demonstrate the applicability of the meta-stereomicroscope to biological imaging, pollen grains with feature sizes ranging from 18 μm to 30 μm are used as test specimens. Their naturally varying surface morphology and depth distribution make them ideal specimens for stereoscopic visualization. The images captured by the left and right optical paths of the meta-stereomicroscope are shown in **Fig. 3(e) and (f)**. Fine structural features of the pollen grains are clearly resolved, confirming the high imaging resolution of the meta-stereomicroscope. Details of the image-processing procedure are provided in **Section S6, Supplemental Information.**

### 3D Surface Morphological Reconstruction

Apart from stereoscopic imaging, the stereoscopic information can in principle be utilized to reconstruct the 3D surface morphology. Depth reconstruction from parallax images is based on corresponding image points (CIPs), defined as matching points of the same object in the left and right views. Larger parallax corresponds to greater depth variation. In microscopic imaging, conventional stereo-matching algorithms have limited ability to extract CIPs from low-detail, low-texture images, thereby hindering accurate depth reconstruction.

With the rapid development of artificial intelligence, transformer-based neural networks with semantic recognition ability provide an effective approach to improving CIPs extraction accuracy under limited data conditions. Here, we develop a stereo neural network for high-accuracy 3D surface morphological reconstruction. A comparison with conventional stereo-matching algorithms is provided in **Section S7, Supplemental Information**.

The network consists of two subnetworks. The first is a stereo-matching network based on the DLNR (Stereo Matching Network with Decouple LSTM (Long Short-Term Memory) and Normalization Refinement) architecture **[36]**, which takes a pair of parallax images as input and estimates the corresponding depth information. The second is a mask network based on the U-Net architecture **[37]**, which extracts the effective pollen grain region and suppresses background misclassification in the stereo-matching process. The final 3D surface morphology is reconstructed by combining the outputs of these two subnetworks. The schematic workflow of the stereo neural network is shown in **Fig. 4**.

Because no publicly available depth datasets exist for microscopic stereoscopic imaging and experimental acquisition of such datasets is challenging, we generated a synthetic training dataset using the open-source rendering software “Blender” from the Blender Foundation. The depth information generated by Blender is referenced to the rendering coordinate system and is therefore calibrated to real-world distances using an electrically driven microscope (Zeiss Axio Observer 7). The training dataset consists of 7,000 pairs of parallax images and corresponding depth information. Network training is performed on an NVIDIA GeForce RTX 3090 Ti GPU.

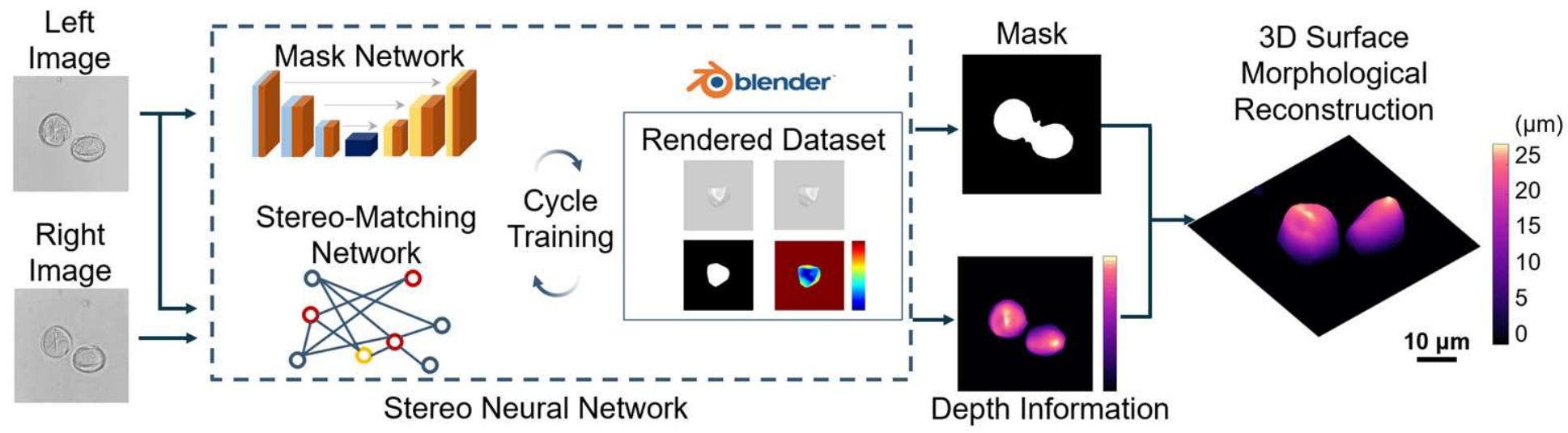

**Fig. 4. Schematic workflow of the stereo neural network.** The network consists of two subnetworks. The mask network and the stereo-matching network are trained using a Blender-rendered dataset with depth information. Left and right images are input to the two subnetworks to generate a mask and depth information, which are then fused to produce the final 3D surface morphological reconstruction.

The right panel in **Fig. 4** presents the 3D surface morphological reconstruction of pollen grains generated by the stereo neural network using **Fig. 3(e)** and **(f)**, enabling intuitive visualization of surface morphology and depth information. Three different pollen grain samples are used for calibration and another eight different pollen grain samples for validation. The resulting depth precision is approximately 730 nm, quantified by the root mean square error (RMSE) between the calibrated depth predictions of the neural network and the ground-truth sample depths (see **Section S8, Supplemental Information** for details). Depth resolution is defined as the minimum distinguishable vertical distance between two planes. To characterize the depth resolution, a linear motorized stage (Thorlabs, Z825B) is used to translate the resolution target along the z-axis. The measured depth resolution is 1026 nm (see **Section S9 of the Supplemental Information**). The meta-stereomicroscope represents more than a twofold precision improvement over conventional stereoscopic systems whose depth precision is worse than 1800 nm (see **Section S10, Supplemental Information** for comparison). **Fig. 3(g)** shows the merged result of the 3D surface morphological reconstruction in **Fig. 4** and the bright-field image of the pollen grains in **Fig. 3(e)**, demonstrating the ability to simultaneously achieve high lateral and depth resolution imaging.

The meta-stereomicroscope also enables real-time imaging and 3D surface morphological reconstruction. The stereo neural network performs 3D surface morphological reconstruction at a frame rate of 8.5 fps, whereas the overall reconstruction rate is 5.3 fps, limited by the readout speed of the polarization sensor. To demonstrate this capability, Saccharomyces fungi, with sizes ranging from 2 μm to 6 μm, are used as imaging targets. A microfluidic channel is formed between a microscope slide and a coverslip, allowing the fungi to move with the flowing water. The complete video is provided as **Movie S1**.

## Meta-Stereomicroscope in Reflection Mode

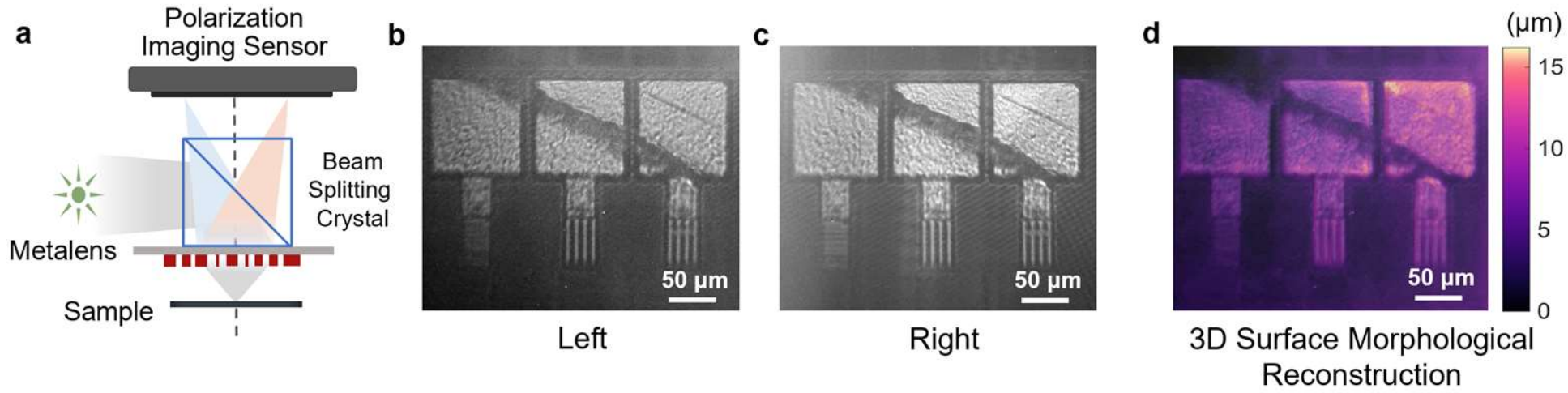

**Fig. 5. Imaging performance of the meta-stereomicroscope under reflection-mode illumination.** (**a**) Schematic of the optical setup. (**b–c**) Imaging of a semiconductor chip from the (**b**) left and (**c**) right optical paths. (**d**) 3D surface morphological reconstruction generated from (**b**) and (**c**) using the stereo neural network.

The meta-stereomicroscope can also operate in reflection mode, which is particularly applicable for imaging opaque samples. The birefringent metalens exhibits unique advantages in reducing reflected background light (see **Section S11 of the Supplemental Information**). To demonstrate this capability, a coaxial reflection-mode meta-stereomicroscope is implemented using a birefringent metalens with a diameter of 2000 μm following the same design principle. It is worth noting that high-performance visible-wavelength metalenses typically rely on meta-atoms with feature sizes below 50 nm, which are challenging to fabricate using scalable and cost-effective commercial manufacturing processes. Here, we explore a more scalable fabrication strategy by employing a reduced-resolution EBL process with relaxed feature-size requirements. The fabricated metalens achieves a minimum feature size of 70 nm (see **Section S12 of the Supplemental Information**), which is comparable to that achieved by roll-to-roll nanoimprint lithography **[38]**, in which an 80 nm feature size has been reported. This result demonstrates the potential compatibility of our design with large-scale manufacturing technologies. The corresponding measured PSFs and MTFs of the 2000 μm-diameter metalens are shown in **Fig. S13(a) and (b)**, from which the maximum resolvable spatial frequencies are 525 lp/mm and 782 lp/mm at an MTF threshold of 0.1 for x- and y-polarized light. The DOF (see **Fig. S13(c)**) of the 2000 μm-diameter metalens is measured to be 7.9 μm, and the focusing efficiencies of the left and right optical paths are 6% and 18%, respectively.

As shown in **Fig. 5(a)**, focused illumination is laterally coupled through a beam-splitting crystal, transmitted through the metalens, and directed onto the sample. The large-aperture metalens facilitates this optical configuration and is advantageous for reflection-mode imaging. The imaging magnification is reduced to 12. Under these conditions, the maximum resolvable spatial frequency at the sensor plane is 840 lp/mm. Although the lower magnification reduces the maximum resolvable spatial frequency, it helps preserve a larger FOV on the sensor. Using a 1951 USAF resolution target, the measured lateral resolutions are 977 nm (512 lp/mm) and 775 nm (645 lp/mm) for the left and right optical paths, respectively, with an FOV diameter of 110 μm (see **Section S14, Supplemental Information**).

A semiconductor chip is used as the imaging target. **Fig. 5(b)** and **5(c)** show the left- and right-view images captured by the reflection-mode meta-stereomicroscope. The relatively high reflectivity of the sample provides a strong signal-to-noise ratio, facilitating high-quality reflection-mode imaging. **Fig. 5(d)** presents the corresponding 3D surface morphological reconstruction generated by the stereo neural network. Fine surface features, including scratches on the chip, are clearly resolved in the reconstructed morphology. The reflection-mode meta-stereomicroscope achieves a depth precision of 1050 nm (see **Section S15** of the **Supplemental Information** for details). The lower magnification also reduces the depth resolution. The experimentally measured depth resolution is 2044 nm (see **Section S16,**

**Supplemental Information**). These results demonstrate its capability for high-precision 3D surface morphological reconstruction of opaque samples.

## 3 Discussion and Conclusion

In conclusion, we demonstrate a fully meta-optical architecture that fundamentally eliminates the FOV mismatch between the metalens and conventional optical components in stereomicroscopes. The demonstrated meta-stereomicroscope is based on a single birefringent metalens that encodes two distinct phase profiles, enabling the simultaneous acquisition of left- and right-view images through orthogonal polarization states. This single device provides flexible control of NA and magnification without requiring optical-axis alignment among multiple components, thereby enabling the simultaneous achievement of high lateral resolution of 435 nm and depth resolution of 1026 nm. Compared with the metalens-assisted stereomicroscope combining a metalens and a commercial stereomicroscope, the demonstrated meta-stereomicroscope eliminates the fundamental FOV mismatch and achieves a larger FOV diameter of 43 μm, compared with that of 6.2 μm achieved by the metalens-assisted stereomicroscope. In addition, the DOF is increased to 8.1 μm, compared with 3.3 μm in the metalens-assisted stereomicroscope, thereby increasing the overall depth measurement range (see **Section S17, Supplemental Information**).

Experimentally, we demonstrate high-resolution stereoscopic imaging in both transmission and reflection modes, achieving performance that surpasses previously reported stereomicroscopes. Combined with the stereo neural network, the meta-stereomicroscope further enables straightforward reconstruction of high-resolution 3D surface morphology. These results establish a compact and highly integrated stereomicroscopy platform with broad potential applications ranging from biomedical imaging to industrial inspection. However, the polarization imaging sensor is costly, and on-chip polarizers restrict the sensor's spatial frequency. Future improvements in metalens design and fabrication may enable operation with conventional image sensors and achromatic imaging capability, further reducing system size while enhancing imaging performance. In addition, due to the inherent limitations of stereo vision, only the depth of the visible surface can be reconstructed. Integrating sample micro-rotation techniques will enable reconstruction of the full 3D surface **[39]**.

## 4 Materials and Methods

### *4.1 Numerical Simulations*

The 3D FDTD method from Ansys is used to construct the meta-atom library and evaluate the focusing characteristics of the birefringent metalens. In the simulation, the complex refractive index of c-Si is used to model the polarization-dependent rectangular meta-atoms at a wavelength of 532 nm. By adjusting the length and the width of the nanopillars, the effective refractive index of the nanopillars could be tailored, enabling a full $2\pi$ phase coverage at a wavelength of 532 nm, as detailed in **Fig. S1** of the **Supplemental Information**. Based on the resulting phase library, meta-atoms are organized according to the phase profile depicted in **Fig. 2(b)**.

### *4.2 Metalens Fabrication*

The metalens is fabricated through a combination of wafer bonding techniques **[40, 41]** and electron beam lithography. The process starts with a silicon-on-insulator (SOI) wafer consisting of a 220-nm-thick c-Si device layer atop a 375-nm silicon dioxide layer. An HSQ resist layer with a thickness of 160 nm is first spin-coated onto the wafer, followed by pattern definition using electron beam lithography (Raith Vistec EBPG-5000 plus ES). After exposure, the designed patterns are transferred into the c-Si layer via inductively coupled plasma (ICP) etching (Oxford Instruments Plasma Pro System 100 ICP180). Subsequently, an additional HSQ layer is deposited and hard-baked to serve as a protective layer. Owing to the excellent backfilling capability of HSQ **[42]**, the resist effectively infiltrates the narrow gaps between adjacent meta-atoms, ensuring structural integrity during subsequent processing. The patterned sample is then bonded to a glass substrate using an ultraviolet (UV)-curable adhesive, Norland Optical Adhesive 61 (NOA61). Finally, the opaque handle substrate is removed through a combination of precision mechanical grinding (Logitech PM6) and ICP etching. A SEM image of the fabricated metalens is shown in the inset of **Fig. 1(c)**.

### *4.3 Optical Characterization*

The imaging performance of the birefringent metalens under transmission-mode illumination is characterized using the optical setup in **Fig. S3(a).** A white-light LED (Thorlabs, MINTL5) equipped with a 532 nm, 10 nm-bandwidth filter (LBTEK, MBF10-532-10) is used as an incoherent, unpolarized transmissive green illumination source. To enhance the imaging resolution, an infinity-corrected objective (20×, NA = 0.4) is used to provide focused illumination. X- and y-polarized parallax images are captured simultaneously by a polarization sensor with on-chip polarizers (Lucid Vision Labs PHX050S1-P/Q).

The optical setup under reflection-mode illumination is shown in **Fig. 5(a)**. The same white-light LED and filter combined with a focusing lens are used to generate a focused illumination beam, and the beam is reflected by a beam-splitting crystal (LBTEK, BS1155-A), transmitted through the metalens to illuminate the sample in reflection mode. The light reflected from the sample is collected by the metalens and imaged onto the polarization sensor.

## 5 Back matter

### *5.1 Funding*

This work is supported by the National Key R&D Program of China (No. 2022YFA1404304), National Natural Science Foundation of China (No. 12374363), Guangdong Provincial Natural Science Fund Projects (No. 2024B1515040013) and Guangdong Provincial Quantum Science Strategic Initiative (GDZX2306002, GDZX2206001). T.F.K. acknowledges support by UKRI contract EP/X037770/1. E.R.M. acknowledges support by CNPq grant 303820/2024-1.

### *5.2 Acknowledgements*

Prof. Martins passed away before the completion of this manuscript. We dedicate this work to his memory.

### *5.3 Disclosures*

The authors declare no conflicts of interest.

### *5.4 Data availability statement*

Data underlying the results presented in this paper are not publicly available at this time but may be obtained from the authors upon reasonable request.

### *5.5 Supplementary Document*

See Supplement 1 and Movie S1 for supporting content.